\documentclass[prl,reprint,nofootinbib,preprintnumbers,amsmath,amssymb]{revtex4}
\usepackage{graphicx,tikz,braket,booktabs}

\usepackage{appendix}
\newcommand{\beq}{\begin{equation}}
\newcommand{\eeq}{\end{equation}}
\newcommand{\bea}{\begin{eqnarray}}
\newcommand{\eea}{\end{eqnarray}}

\def\cN{\mathcal{N}}

\def\cM{\mathcal{M}}

\newcommand{\be}{\begin{equation}}
\newcommand{\ee}{\end{equation}}

\renewcommand{\b}{\beta}

\def\cM{{\cal M}}
\def\cN{{\cal N}}

\def\cO{{\cal O}}

\def\cN{{\cal N}}
\def\cU{{\cal U}}

\begin{document}

\title{\bf Renormalization and running in the 2D $CP(1)$ model }

\medskip\

\medskip\

\author{Diego Buccio${}^1$}
\email{dbuccio@sissa.it}
\author{John F. Donoghue${}^{2}$}
\email{donoghue@physics.umass.edu}
\author{Gabriel Menezes${}^{3}$~\footnote{On leave of absence from Departamento de F\'{i}sica, Universidade Federal Rural do Rio de Janeiro.}}
\email{gabriel.menezes10@unesp.br}
\author{Roberto Percacci${}^1$}
\email{percacci@sissa.it}
\affiliation{
${}^1$International School for Advanced Studies,
Via Bonomea 265, 34134 Trieste, Italy
and INFN, Sezione di Trieste, Trieste, Italy\\
${}^2$Department of Physics,
University of Massachusetts,
Amherst, MA  01003, USA\\
${}^3$Instituto de F\'isica Te\'orica, Universidade Estadual Paulista,
Rua Dr.~Bento Teobaldo Ferraz, 271 - Bloco II, 01140-070 S\~ao Paulo, S\~ao Paulo, Brazil}

\begin{abstract}
We calculate the scattering amplitude in the two dimensional $CP(1)$ model in a regularization scheme independent way. When using cutoff regularization, a new Feynman rule from the path integral measure is required if one is to preserve the symmetry. The physical running of the coupling with renormalization scale arises from a UV finite Feynman integral in all schemes.  We reproduce the usual result with asymptotic freedom,  but the pathway to obtaining the beta function can be different in different schemes. We also comment on the way that this model evades the classic argument by Landau against asymptotic freedom in non-gauge theories.
\end{abstract}

\maketitle

\section{1. Introduction and summary}

The two dimensional $CP(1)$ nonlinear sigma model~\footnote{A few words on terminology. Historically, the $CP(1)$ model was formulated with two complex scalars and a U(1) gauge invariance. Indeed, early references for CP$(N-1)$ models talk about this formulation~\cite{Eichenherr:1978qa,Golo:1978de,Cremmer:1978bh,Bardeen:1976zh,Brezin:1976qa,DAdda:1978vbw,DAdda:1978dle,Din:1979zv,Novikov:1984ac}. In this work we will use the formulation as given in Ref.~\cite{Shifman:2012zz}. In the end such formulations must be equivalent because $CP(1)$$=S^2$. Indeed, what we do here -- the sphere in stereographic coordinates -- can legitimately
be called $CP(1)$ model, but long before that it was also referred to as $O(3)$ model~\cite{Polyakov:1975yp}.} is defined by the Lagrangian
\beq\label{CP1}
{\cal L} = \frac{\partial_\mu \phi^* \partial^\mu \phi}{(1+\frac{g^2}{2} \phi^*\phi)^2}\ .
\eeq
Despite its nonlinear nature, with the Lagrangian containing all powers of the fields, it is renormalizable. It is also interesting because the coupling is asymptotically free. 

Our focus here is the regularization scheme dependence. Present calculations have been done using cutoffs, with the tadpole loop diagram playing a prominent role in obtaining asymptotic freedom. However, the tadpole diagram vanishes in dimensional regularization, which initially appears puzzling. Moreover, the tadpole diagram does not contain any momentum dependence, so it is not initially clear how the scattering amplitudes of the theory would manifest asymptotic freedom in their kinematic dependence. Therefore we will calculate a scattering amplitude in various regularization schemes.  The result does have some interesting features. Because the calculation itself is relatively simple, we will use this introduction also as a summary of the main features of the result.

When naively using a cutoff regularization, one finds  loop effects which violate the symmetry of the theory. For example, with the scattering diagrams shown in Figure 1, the tadpole diagram is obtained from the interaction $\frac{3g^4}{4}\partial_\mu\phi^* \partial^\mu\phi (\phi^*\phi)^2$. The contraction of the two fields involving derivatives leaves behind an interaction with no factors of momentum which is different from any term in the original Lagrangian. Moreover it involves the quadratically divergent integral
\beq\label{quadratic} 
\int^\Lambda \frac{d^2p}{(2\pi)^2}  \frac{p^2}{p^2+i\epsilon} \sim \Lambda^2
\eeq 
In Section 2 we describe the origin of Feynman rules following from the path integral measure which have the effect of removing these symmetry violating quadratic divergences from the theory. This modification is not required in dimensional regularization because the scaleless integral of Eq. \ref{quadratic} is set equal to zero.

When renormalizing the theory one is confronted with both UV divergences and IR divergences. Somewhat surprisingly all 2D sigma models with $O(N)$ symmetry, including the $CP(1)$ model, are known to be IR finite \cite{David:1980rr}. In Sec. 3, we use the background field method to renormalize the 4-point vertex of this theory and to verify the IR finiteness.

\usetikzlibrary {arrows.meta}
\begin{figure}[ht]
\begin{center}
\begin{tikzpicture}
[>=stealth,scale=0.66,baseline=-0.1cm]
\draw (-1,0) arc (180:0:1cm);
\draw (-1,0) arc (180:360:1cm);
\draw [] (-1,0) node[anchor=west] {} -- (-2.4,1);
\draw [] (-1,0) node[anchor=west] {} -- (-2.4,-1);
\draw [] (2.4,1) -- (1,0) node[anchor=east] {};
\draw [] (2.4,-1) -- (1,0) node[anchor=east] {};
\draw[fill] (1,0) circle [radius=0.1];
\draw[fill] (-1,0) circle [radius=1mm];
\end{tikzpicture}
\qquad
\begin{tikzpicture}
[>=stealth,scale=0.66,baseline=-0.1cm]
\draw (-1,0) arc (180:0:1cm);
\draw (-1,0) arc (180:360:1cm);
\draw [](2,-1) node[anchor=west] {} -- (0,-1);
\draw [](0,-1) -- (-2,-1) node[anchor=east] {};
\draw [](0,-1) -- (-2,-2) node[anchor=east] {};
\draw [](0,-1) -- (2,-2) node[anchor=east] {};
\draw[fill] (0,-1) circle [radius=1mm];
\end{tikzpicture}
\caption{Diagrams associated with one-loop scattering.}
\label{fig:feynman}
\end{center}
\end{figure}
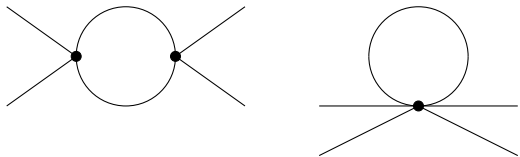

The calculation of the amplitude is given in Sec 4. We use the Passarino-Veltman reduction method to  calculate the scattering amplitude for the reaction 
$\phi_1 +\phi_1\to \phi_2+\phi_2$ using real fields defined by 
$\phi = (\phi_1+i\phi_2)/\sqrt 2$. In any renormalization scheme we find 
\bea\label{amplitude}
{\cal M} &=& g_0^2 s -\frac{g_0^4s}{4}\left[I(t)+I(u)\right] \nonumber \\
&~&~~~~~ + \frac{g_0^4}{4}(u-t)[I(t)-I(u)]
\eea
where $I(q^2)$ is a specific combination of the scalar tadpole and scalar bubble diagrams
\beq
I(q^2) = 2T - q^2 B(q^2)
\label{irfinite}
\eeq
with the two dimensional tadpole and bubble diagrams defined by 
\bea\label{tadpolebubble}
T  &=&-i  \int \frac{d^2p}{(2\pi)^2} \frac1{p^2+i\epsilon}  \nonumber \\
B(q^2) &=& -i \int \frac{d^2p}{(2\pi)^2} \frac1{[p^2+i\epsilon][(p-q)^2+i\epsilon]}\ .
\eea
The combination $I(q^2)$ is the unique combination of the scalar tadpole and the scalar bubble diagrams which is IR finite. Given the IR finiteness of all $O(N)$ sigma models, this must be the combination which appears in all amplitude calculations. 

If we define the coupling at the renormalization point $s=t=u=\mu_R^2$, we can define the renormalized coupling as
\beq
g_r^2(\mu_R) = g_0^2 - \frac{g_0^4}{2} I(\mu^2_R)  \ \ .
\eeq
In any renormalization scheme we have 
\beq
I(q^2) - I(\mu_R^2) = \frac1{2\pi}\log (\mu_R^2/q^2)
\eeq
so that the renormalized amplitude becomes
\bea
{\cal M} &= &g_r^2(\mu_R^2) s 
-\frac{g_r^4 \, s}{8\pi} \left(\log (-t/\mu_R^2)  +\log(-u/\mu_R^2) \right))
-\frac{g_r^4}{8\pi} (t-u)\log (t/u)\ .
\eea
This form implies the beta function
\beq
\beta_g= \mu_R \frac{\partial g_R(\mu_R)}{\partial \mu_R} = -\frac{g^3}{4\pi}
\eeq
which is the usual answer implying asymptotic freedom. 

While this result is satisfying and perhaps not surprising, the way that this emerges in specific schemes is remarkably different. We here discuss three possible schemes.
\\\\\\\\\\\\\\\\\\\\\\\\\\\\\\\\\\\\\\\\\\\
i) Cutoff regularization. In this scheme the tadpole diagram has both UV and IR sensitivity
\beq
T =- \frac1{4\pi} \log\frac{\Lambda^2}{k^2}
\eeq
where $\Lambda$ is a UV cutoff and $k$ is an IR cutoff. The scalar bubble diagram is UV finite
\beq
q^2 B(q^2) = -\frac1{2\pi} \log\frac{-q^2}{k^2}
\eeq
\\
ii) Dimensional regularization. Here the tadpole diagram is scaleless and hence vanishes
\beq
T=0
\eeq
The scalar bubble diagram is again UV finite but has in IR divergence, with the result
\beq
q^2B(q^2) = \frac1{2\pi} \left[ \frac1{\epsilon} - \log\frac{-q^2}{\mu^2}  \right]
\eeq
\\
with $\epsilon = (2-d)/2$.
\\
iii) Hybrid regularization. Here we use an IR cutoff $k$, but use dimensional regularization in the UV. In this case
\beq
T =- \frac1{4\pi} \left[ \frac1{\epsilon} - \log\frac{k^2}{\mu^2}  \right]
\eeq
while the bubble diagram is the same as in the cutoff regularization
\beq
q^2 B(q^2) =- \frac1{2\pi} \log\frac{-q^2}{k^2}\ .
\eeq

In calculations in the literature (see e.g. \cite{Shifman:2012zz}), the running has been calculated using cutoff regularization by following the UV cutoff $\Lambda$. 
We can call this cutoff running. 
In the amplitude for the process
$\phi_1 +\phi_1\to \phi_2+\phi_2$
the bubble is finite and $\Lambda$ appears uniquely in the tadpole diagram.  In contrast, in dimensional regularization the tadpole diagrams vanish. The UV divergence of the tadpole is replaced by a factor of $1/\epsilon$ of IR origin in the bubble diagram, with the accompanying factor of $\log \mu$. In the hybrid scheme, the UV divergence of $1/\epsilon$ reappears in the tadpole diagram.
In these cases, one can follow the running by following $\log \mu$, which can be called $\mu$ running. However, since both $\Lambda$ and $\mu$ disappear from renormalized amplitudes, the running that is observed in practice involves the behavior of the amplitudes with the kinematic variables. We call this physical running. We have shown how all schemes lead to the same physical running.

For this theory, the equivalence of all schemes is basically due to dimensional analysis. There are no explicit dimensional factors in the Lagrangian, and it is also important that there is no sensitivity to a potential infrared cutoff. Therefore the logarithms must involve $\log (\Lambda^2/E^2) $ or $\log (\mu^2/E^2) $ (where $E$ is a common energy scale in the amplitude), and the $\Lambda$ or $\mu$ behavior tracks the physical running. However there are other theories with explicit dimensionful parameters where this is no longer true. Our other recent papers \cite{Buccio:2023lzo,Donoghue:2023yjt,Buccio:2024hys} provide examples of this. We return to this aspect in the closing discussion of Section 5. 

A somewhat surprising aspect of this result is that the running of the amplitude with the physical momentum comes from the UV finite bubble diagram in all schemes. In four dimensions with mass-independent renormalization, we are used to having the bubble diagram provide both the UV divergence and the related factor of $\log E^2$ which provides the physical running. In two dimensions the bubble is still the only source of the dependence on the physical momenta, but it is UV finite and IR sensitive. A focus only on the divergent diagrams may be dangerous unless a full calculation is performed. We have found this lesson also in our treatment of physical running in Quadratic Gravity \cite{Buccio:2024hys}. 
 
In the final discussion, we also comment on the existence of asymptotic freedom in a non-gauge theory.

\section{2. Regularization scheme and Feynman rules}

In theories with Goldstone bosons, the symmetry requires that the interaction terms involve derivatives. This forbids mass terms and also enforces the Adler zeros for scattering amplitudes. However in such theories, there can be loops where the derivatives act on the loop particles leaving no momentum factors for the external particles. Besides the example given in the introduction, the $CP(1)$ theory could generate a mass term from the interaction $\partial_\mu \phi^*\partial^\mu \phi( \phi^*\phi)$ if the two fields with derivatives are contracted in a tadpole loop. This problem was known for sigma models in the 1960's and the resolution is a contribution to the Feynman rules from the path integral measure, which cancels the offending diagrams and preserves the symmetry\cite{Honerkamp:1996va, Gerstein:1971fm}. For our purposes the simplest way to obtain this factor is to follow the derivation of the Lagrangian of Eq. 1 from the constrained $O(3)$ version. The $CP(1)$ model describes a spin system in 2D 
\beq
S = \frac1{2g^2} \int d^2 x~\partial_\mu \mathbf{S}(x) \cdot \partial^\mu \mathbf{S}(x)
\eeq
with $\mathbf{S}= (S_1, S_2, S_3)$, subject to the constraint
\beq
\mathbf{S}(x)\cdot \mathbf{S}(x) =1 \ \ .
\eeq
In order to change to the unconstrained version, one identifies
\bea
S_1 &=& \frac{g\phi_1}{1+\frac{g^2}{4} (\phi_1^2 +\phi_2^2)}  \nonumber \\
S_2 &=& \frac{g\phi_2}{1+\frac{g^2}{4} (\phi_1^2 +\phi_2^2)}  \nonumber \\
S_3 &=& \frac{1-\frac{g^2}{4} (\phi_1^2+\phi^2_2)}{1+\frac{g^2}{4} (\phi_1^2 +\phi_2^2)}   
\eea
which reproduces the Lagrangian of Eq.  \ref{CP1} with $\phi= (\phi_1+i\phi_2)/\sqrt{2}$.  However, the transformation has a Jacobian. Following Honerkamp and Meetz \cite{Honerkamp:1996va} (see also \cite{Gerstein:1971fm,  DeWitt:1962ud, Boulware:1970zc, Salam:1971sp, Charap:1970xj, Donoghue:2020hoh}) we are lead to the  invariant measure 
\beq
\int \frac{[d\phi]}{1+\frac{g^2}{2}\phi^*\phi}
\eeq
This can be converted into an interaction  by exponentiation
\beq
e^{\delta^2 (0)\int d^2x \log (1+\frac{g^2}{2} \phi^*\phi ) }
\eeq
The leading $\delta^2(0) $ vanishes in dimensional regularization because it is a scaleless integral. However, in cutoff regularization this cancels the $\Lambda^2 $ contributions found in the Feynman diagrams, preserving the symmetry. Note that one can also obtain the invariant measure in a somewhat more cumbersome way using canonical quantization through a modification in the canonical momenta in theories with higher derivatives \cite{Gerstein:1971fm}.

\section{3. Background field renormalization and IR finiteness}

  In this section we use a hybrid renormalization scheme to renormalize the four-point vertex in the background field method. This will show the infrared finiteness of the vertex and we will obtain the running coupling using what we refer to as $\mu$ running. The renormalization scheme uses an infrared cutoff during intermediate steps, and regularizes the UV physics with dimensional regularization. This feature permits to easily distinguish the two types of divergences.

Let's consider a generic NLSM in Euclidean space of dimension $d=2$
\be
S[\varphi]=\frac{1}{2g^2} \int d^2x\, h(\varphi)_{\alpha\beta}
\partial_\mu\varphi^\alpha\partial^\mu\varphi^\beta\label{action}
\ee
The background field method applied to NLSM has been widely discussed in literature \cite{Honerkamp:1971sh,Alvarez-Gaume:1981exa,Howe:1986vm}. The second variation of the action can be put in a covariant form with respect to coordinate transformation in the target space using a nonlinear split. The quantum field $\xi^\alpha(x)$ is a vector field,
related to the total field $\varphi^\alpha(x)$ and the background field $\bar{\varphi}^\alpha(x)$ by the exponential map
\be
\varphi(x)=\exp_{\bar{\varphi}(x)}(\xi(x))\ .
\ee
With this choice, the action up to order $\xi^2$ is
\be
S[\bar{\varphi}]
+\frac{1}{g^2}\int d^2x\,
h(\bar{\varphi})_{\alpha\beta}\partial_\mu\bar{\varphi}^\alpha D_\mu\xi^\beta
+\frac{1}{2g^2}\int d^2x\,\xi^\alpha\left(
-h(\bar{\varphi})_{\alpha\beta}D_\mu D^\mu
+\partial_\mu\bar{\varphi}^\gamma\partial^\mu\bar{\varphi}^\delta R_{\alpha\gamma\beta\delta}\right)\xi^\beta\label{action2}
\ee
where the covariant derivative $D_\mu\xi^\alpha$ is defined as
\be
D_\mu\xi^\alpha=\partial_\mu\xi^\alpha+\partial_\mu\bar{\varphi}^\gamma\Gamma_\gamma{}^\alpha{}_\beta \xi^\beta
\ee
and $\Gamma$ and $R$ are respectively the Christoffel symbols and the Riemann tensor of the metric $h_{\alpha\beta}(\bar{\varphi})$ on the target space.
The linear term gives the classical equations of motion
\be
\partial_\mu\partial^\mu\varphi^\alpha+\partial_\mu\varphi^\gamma\partial^\mu\varphi^\beta\Gamma_\gamma{}^\alpha{}_\beta=0 \label{eom}
\ee
while the quadratic term can be written as $\frac{1}{2g^2}\int\xi^\alpha \mathcal{O}_{\alpha\beta}\xi^\beta$ and used to compute the one-loop correction to the quantum effective action
\be
\Gamma[\bar\phi]=S[\bar\phi]+\frac12 \mathrm{tr}\log\cO\ .
\ee
The propagator in (\ref{action2}) has a nonstandard form due to the metric $h_{\alpha\beta}(\bar{\varphi})$, that forbids an easy Fourier transform and diagrammatic computation. A way around consists in defining the vielbein $e^a{}_\alpha$, the vector field $\xi^a=\xi^\alpha e^a{}_\alpha$ and the spin connection $\omega_\gamma{}^{ab}$. The new covariant derivative is
\be
D_\mu\xi^a=\partial_\mu\xi^a+\partial_\mu\bar{\varphi}^\gamma\omega_\gamma{}^a{}_b \xi^b
\ee
and the second order term in the $\xi$ expansion of the action turns into
\be
S^{(2)}= \frac{1}{2g^2}\int d^2x\,  \xi^a\left( -D_\mu D^\mu \delta_{ab}
+\partial_\mu\bar{\varphi}^\gamma\partial_\mu\bar{\varphi}^\delta R_{a\gamma b\delta}\right)\xi^b
\ee
with $R_{a\gamma b\delta}=R_{\alpha\gamma\beta\delta}e_a{}^\alpha e_b{}^\beta$.
In this way $\cO$ can be written in the form
\be
\cO_{ab}=\delta_{ab}\partial_\mu\partial^\mu+\cN^\mu_{ab}\partial_\mu+\cU_{ab}\ .\label{O}
\ee
If $\cN$ and $\cU$ are small with respect to the kinetic term, one can perturbatively expand $\mathrm{tr}\log\cO$ and find
\be
\frac12 \mathrm{tr}\log\cO=\frac12\mathrm{tr}\left[\frac{\cU}{\Box}-\frac12\left(\frac{1}{\Box}\cU\frac{1}{\Box}\cU+\frac{1}{\Box}\cN\partial\frac{1}{\Box}\cN\partial+2\frac{1}{\Box}\cU\frac{1}{\Box}\cN\partial\right)\right]\label{1loopG}
\ee
where, in the hybrid renormalization scheme,
\bea
\mathrm{tr}\frac{\cU}{\Box}&=&-\frac{1}{4\pi}\cU_{aa}\left[-\frac1\epsilon+\log\left(\frac{\mu^2}{m^2}\right)\right]\\
\mathrm{tr}\frac{1}{\Box}\cU\frac{1}{\Box}\cU&=&-\frac{1}{4\pi}(\cU_{ab}+\cU_{ba})\log\left(-\frac{\Box}{m^2}\right)\frac{1}{\Box}\cU^{ab}
\\
\mathrm{tr}\frac{1}{\Box}\cU\frac{1}{\Box}\cN\partial &=&-\frac{1}{4\pi}(\cU_{ab}+\cU_{ba})\log\left(-\frac{\Box}{m^2}\right)\frac{1}{\Box}\partial_\mu\cN^{\mu ab}
\\
\mathrm{tr}\frac{1}{\Box}\cN\partial\frac{1}{\Box}\cN\partial &=&\frac{1}{16\pi}\cN_{\mu ab}\left(-\frac1\epsilon-\log\left(-\frac{\Box}{\mu^2}\right)\right)\left(\cN^{\mu ab}-\cN^{\mu ba}\right)\nonumber
\\
&&+\frac{1}{8\pi}\cN_{\mu ab}\log\left(-\frac{\Box}{m^2}\right)\frac{\partial_\mu\partial_\nu}{\Box}\cN^{\nu ab}
\eea
up to finite terms.
$\cU$ is symmetric in vielbein indices, while $\cN$ is antisymmetric, hence the $\cU\cN$ bubble is identically zero.

In the particular case of the $CP(1)$ nonlinear sigma model the metric is simply 
$$h_{\alpha\beta}=\frac{\delta_{\alpha\beta}}{[1+\frac14(\varphi^\alpha\varphi^\alpha)]^2}
$$
with indices running from 1 to 2. The curvatures tensors and connections can be computed  using standard differential geometry. Notice that with a simple rescaling $\phi=\frac1g \varphi$ we get exactly the action (\ref{CP1}).

Since we are interested in the one loop correction to the four point function, we can use the small $\phi$ expansion 
\be
S[\phi]=\frac{1}{2} \int d^2x\, \delta_{\alpha\beta} \partial_\mu\phi^\alpha\partial^\mu\phi^\beta\left[1-\frac{g^2}{2}\phi^\gamma\phi^\gamma+\frac{3g^4}{16}(\phi^\gamma\phi^\gamma)^2\right]+O(\phi^8).
\label{CP1pert}
\ee
The classical equations of motion are
\be
\partial_\mu\partial^\mu\phi^\alpha=\frac{g^2}{2}\partial_\mu\phi^\gamma\partial^\mu\phi^\beta(\delta^\alpha_\beta \phi_\gamma+\delta^\alpha_\gamma \phi_\beta-\delta_{\gamma\beta}\phi^\alpha) + O(\phi^5)
\ee
and we just need
\bea
\cN^\mu{}_{ab}&=&-g^2\partial^\mu\bar\phi^\gamma (\delta^a_\gamma\bar\phi^b-\delta^b_\gamma\bar\phi^a)+O(\bar\phi^4)
\\
\cU_{ab}&=&g^2\partial_\mu\bar\phi^\gamma\partial^\mu \bar\phi^\delta\left[(1-\frac{g^2}{2}\bar\phi^\gamma\bar\phi^\gamma)(\delta_{ab}\delta_{\gamma\delta}-\delta_{a\delta}\delta_{\b\gamma})+\frac{g^2}{4}\bar\phi^\alpha\partial_\mu\bar\phi^\alpha (\bar\phi_a\partial^\mu\bar\phi_b+\bar\phi_b\partial^\mu\bar\phi_a)\right.\nonumber
\\
&&\left.-\frac{g^2}{4}\partial_\mu\bar\phi_a \partial^\mu\bar\phi_b \bar\phi^\alpha\bar\phi^\alpha-\frac{g^2}{4}\partial_\mu\bar\phi^\alpha\partial^\mu\bar\phi^\alpha \bar\phi_a \bar\phi_b\right]+O(\bar\phi^6)
\eea
to compute (\ref{1loopG}) at order $\bar\phi^4$.
So the bubble contributions are
\bea
\mathrm{tr}\frac{1}{\Box}\cU\frac{1}{\Box}\cU&\approx&-\frac{g^4}{2\pi}\partial_\mu\bar\phi^\gamma\partial_\mu \bar\phi^\delta\log\left(-\frac{\Box}{m^2}\right)\frac{1}{\Box}\partial_\nu\bar\phi^\gamma\partial_\nu \bar\phi^\delta
\\
\mathrm{tr}\frac{1}{\Box}\cN\partial\frac{1}{\Box}\cN\partial &\approx& \frac{g^4}{4\pi}\left[\bar\phi^\alpha\partial_\mu\bar\phi^\beta\left(-\frac1\epsilon-\log\left(-\frac{\Box}{\mu^2}\right)\right)(\bar\phi^\alpha\partial_\mu\bar\phi^\beta-\bar\phi^\beta\partial_\mu\bar\phi^\alpha)\right]\nonumber
\\
&&+\frac{g^4}{4\pi} \left[\bar\phi^\alpha\partial_\mu\bar\phi^\beta \log\left(-\frac{\Box}{m^2}\right)\frac{\partial_\mu\partial_\nu}{\Box}(\bar\phi^\alpha\partial_\nu\bar\phi^\beta-\bar\phi^\beta\partial_\nu\bar\phi^\alpha)\right]
\eea
In this approximation, thanks to the equations of motions, we can write
\be
\Box(\bar\phi^\alpha\bar\phi^\beta)=2\partial_\mu\bar\phi^\alpha\partial^\mu\bar\phi^\beta+O(\bar\phi^4)
\ee
and the bubble diagrams reduce to
\bea
\mathrm{tr}\frac{1}{\Box}\cU\frac{1}{\Box}\cU&\approx&-\frac{g^4}{4\pi}\bar\phi^\gamma \bar\phi^\delta\log\left(-\frac{\Box}{m^2}\right)\partial_\mu\bar\phi^\gamma\partial_\mu \bar\phi^\delta
\\
\mathrm{tr}\frac{1}{\Box}\cN\partial\frac{1}{\Box}\cN\partial &\approx& \frac{g^4}{4\pi}\left[\bar\phi^\alpha\partial_\mu\bar\phi^\beta\left(-\frac1\epsilon-\log\left(-\frac{\Box}{\mu^2}\right)\right)(\bar\phi^\alpha\partial_\mu\bar\phi^\beta-\bar\phi^\beta\partial_\mu\bar\phi^\alpha)\right]
\eea
By an integration by parts we can move the partial derivative from one side to the other of $\log{(-\Box)}$, obtaining
\bea
\mathrm{tr}\frac{1}{\Box}\cU\frac{1}{\Box}\cU+\mathrm{tr}\frac{1}{\Box}\cN\partial\frac{1}{\Box}\cN\partial\approx \frac{g^4}{2\pi}\bar\phi^\alpha\partial_\mu\bar\phi^\beta\log(-\Box)\bar\phi^\beta\partial_\mu\bar\phi^\alpha+\frac{g^4}{4\pi}\log(m^2) \bar\phi^\alpha\partial_\mu\bar\phi^\alpha \bar\phi^\beta\partial_\mu\bar\phi^\beta\nonumber
\\
\frac{g^4}{4\pi}\left[-\frac1\epsilon+\log\left(\mu^2\right)\right]\left(\bar\phi^\alpha\bar\phi^\alpha \partial_\mu\bar\phi^\beta\partial_\mu\bar\phi^\beta-\bar\phi^\alpha\partial_\mu\bar\phi^\alpha \bar\phi^\beta\partial_\mu\bar\phi^\beta\right)
\eea
On the other hand, the tadpole at order $\bar\phi^4$ is
\bea
\mathrm{tr}\frac{\cU}{\Box}\approx-\frac{g^2}{4\pi}\left[\partial_\mu\bar\phi^\beta\partial^\mu\bar\phi^\beta\left(1-\frac{g^2}{2}\bar\phi^\alpha\bar\phi^\alpha\right)^2+\frac{g^2}{2}(\bar\phi^\alpha\partial_\mu\bar\phi^\alpha)^2\right]\left[-\frac1\epsilon+\log\left(\frac{\mu^2}{m^2}\right)\right]\nonumber
\\
\approx-\frac{g^2}{4\pi}\left[\partial_\mu\bar\phi^\beta\partial^\mu\bar\phi^\beta(1-g^2\bar\phi^\alpha\bar\phi^\alpha)+\frac{g^2}{2}(\bar\phi^\alpha\partial_\mu\bar\phi^\alpha)^2\right]\left[-\frac1\epsilon+\log\left(\frac{\mu^2}{m^2}\right)\right],
\eea
then
\bea
\frac12 \mathrm{tr}\log\cO&\approx&-\frac{g^2}{8\pi}\partial_\mu\bar\phi^\beta\partial^\mu\bar\phi^\beta(1-\frac{g^2}{2}\bar\phi^\alpha\bar\phi^\alpha)\left[-\frac1\epsilon+\log\left(\mu^2\right)\right]\nonumber
\\
&&+\frac{g^2}{8\pi}\partial_\mu\bar\phi^\beta\partial^\mu\bar\phi^\beta (1-g^2\bar\phi^\alpha\bar\phi^\alpha)\log\left(m^2\right)-\frac{g^4}{8\pi}\bar\phi^\alpha\partial_\mu\bar\phi^\beta\log(-\Box)\bar\phi^\beta\partial_\mu\bar\phi^\alpha \label{1loop}
\eea
Again, the equations of motion implies on-shell
\be
\bar\phi^\alpha\partial_\mu\partial^\mu\bar\phi^\alpha\approx -g^2\partial_\mu\bar\phi^\beta\partial^\mu\bar\phi^\beta \bar\phi^\alpha\bar\phi^\alpha\ ,
\ee
so the term proportional to $\log m^2$ in (\ref{1loop}) can be set to zero via integration by parts. That is the clear sign that IR divergences cancel out as expected in the four-point amplitude. At the same time, the rest of the one-loop correction to the effective action reduces to
\be
\frac12 \mathrm{tr}\log\cO\approx -\frac{g^4}{16\pi}\partial_\mu\bar\phi^\beta\partial^\mu\bar\phi^\beta\bar\phi^\alpha\bar\phi^\alpha\left[-\frac1\epsilon+\log\left(\mu^2\right)\right]-\frac{g^4}{8\pi}\bar\phi^\alpha\partial_\mu\bar\phi^\beta\log(-\Box)\bar\phi^\beta\partial_\mu\bar\phi^\alpha 
\ee
Comparing the last expression with (\ref{CP1pert}), one can renormalize the coupling in the following way
\be
g^2(\mu)=g_B^2-\frac{g_B^4}{4\pi }\log\left(\mu^2\right)\ ,
\ee
from which we recover the $\mu$-running
\be
\beta_\mu(g)=-\frac{1}{4\pi }g^3
\ee
The term containing $\log(-\Box)$ generates at tree level the one loop amplitude (\ref{amplitude}), hence the physical running of the four-point function is equal to the $\mu$-running.

From \cite{David:1980rr}, the expectation value of $\mathcal{L}$ should be independent of $\log(m^2)$.
However, the well known expression in the literature for the one loop effective action \cite{Shifman:2012zz}
\be
\mathcal{L}^{(0)}+\mathcal{L}^{1-loop}=2\left[\frac{1}{g_0^2}-\frac{1}{4\pi}\log\left(\frac{\Lambda^2}{m^2}\right)\right] \frac{\partial_\mu \phi^* \partial^\mu \phi}{(1+\phi^*\phi)^2}\label{1LEA}
\ee
clearly depends on $m$.
In the small coupling (or small field) expansion $\langle\mathcal{L}\rangle$ should be independent of $m$ order by order, but the tree level four-point amplitude from (\ref{1LEA}) is
\be
s\left[\frac{1}{g_0^2}-\frac{1}{4\pi}\log\left(\frac{\Lambda^2}{m^2}\right)\right]^{-1}\ .
\ee
The usual solution consists in identifying $m$ with the running scale $\mu$ and forget about the actual origin of $\log(\mu)$ terms. In our calculation we have seen that this substitution is a nontrivial effect due to the presence of non-local operators in the one loop effective action. These terms, generated by the $p\ll q$ regions of Feynman integrals, cancel all $m$ dependence from the four-point amplitude and allow us to reproduce at tree level the correct momentum structure near the renormalization point.

\section{4. Passarino-Veltman reduction and the scattering amplitude}

In two dimensions, the Passarino-Veltman reduction technique says that all one loop integrals can be reduced to momentum dependent factors times the scalar bubble and tadpole diagrams, given in Eq. (\ref{tadpolebubble}).  In calculating the amplitude, we need only one such example, which is the tensor bubble diagram given by
\beq
\label{tensor}
B_{\mu\nu} (q) =-i \int  \frac{d^dp}{(2\pi)^d}  \frac{p_\mu p_\nu}{[p^2+i\epsilon][(p-q)^2+i\epsilon]} = 
\frac{2 T}{4 (d-1)} \left( \eta^{\mu  \nu }
+ \frac{(d-2) q^{\mu } q^{\nu }}{q^2} \right)
- \frac{q^2B(q) }{4(d-1)} \left(\eta_{\mu\nu} - d\frac{q_\mu q_\nu }{q^2}\right)
\eeq
 in any dimension and any renormalization scheme. Only the scalar bubble carries momentum dependence. This reduction makes it simple to calculate the scattering amplitude. 
 
 Let us first give the calculation for $\phi_1+\phi_1 \to \phi_2+\phi_2 $ in pure dimensional regularization. Here the tadpole diagram vanishes because it is scaleless. The relevant amplitudes are
 \bea\label{different}
 -i{\cal M}(\phi_1(p_1)\phi_1(p_2)\to \phi_2(p_3)\phi_2(p_4)) = i g_0^2 [p_1\cdot p_2+p_3\cdot p_4]  \to  ig_0^2 s
 \eea
 and
\bea\label{similar}
 -i{\cal M}(\phi_1(p_1)\phi_1(p_2)\to \phi_1(p_3)\phi_1(p_4)) \to 0
 \eea
where the second form in each case is the on-shell amplitude. When combining these into loop amplitudes of Fig 1, we find that the $s$-channel loop vanishes, while the $t$-channel and $u$-channel loops are non-vanishing
\footnote{Here and throughout this paper we are focusing only on the divergences and the logarithms, and hence drop terms of order $\epsilon$} .  
Using the reduction technique, we find
 \bea
{\cal M} &=& g_0^2 s +\frac{g_0^4s}{4}\left[tB(t)+uB(u)\right] 
- \frac{g_0^4}{4}(u-t)[tB(t)-uB(u)]
\eea
 in terms of the scalar bubble diagram. The renormalization and running of this amplitude is analyzed in the Introduction. We also note that this amplitude can be constructed directly using unitarity based methods, in which the cuts are calculated in all channels and the logarithms are related to the value of the on-shell cut amplitude. The vanishing of the 
$s$-channel loop is directly related to the vanishing of Eq. (\ref{similar}) on-shell.

We can readily convert this result to any renormalization scheme by using the knowledge that the result is independent of an infrared regulator, known from \cite{David:1980rr} and confirmed in the previous section. This is because the only combination of the scalar tadpole and scalar bubble which is IR independent is the combination (\ref{irfinite}).
The complete result is then of the form of Eq. (\ref{amplitude}).

\section{5. Discussion}

  
Here we would like to clarify the meaning of ``physical running''
and to comment on the relation of the calculation in this paper 
to our other recent papers where we have explored the issue of physical running and regularization schemes \cite{Buccio:2023lzo,Donoghue:2023yjt,Buccio:2024hys}. 
Consider a hypothetical amplitude of the form
\be
\cM(p)=\lambda(\mu)+a\lambda^2(\mu)\log\left(\frac{m^2}{\mu^2}\right)
+b\lambda^2(\mu)\log\left(\frac{p^2}{\mu^2}\right)
+c\lambda^2(\mu)\log\left(\frac{p^2}{m^2}\right)\ ,
\ee
where $\mu$ comes from dimensional regularization
and $m$ is either a mass that is present in the theory
or an IR regulator.
Such an amplitude could arise, for example,
from application of the $\overline{\mathrm{MS}}$ scheme.
The first term represents the contribution of the tadpole,
the other two are the contributions of the UV and IR  part of the bubble.
We can also rewrite the amplitude as
\be
\cM(p)=\lambda(\mu)+(a-c)\lambda^2(\mu)\log\left(\frac{m^2}{\mu^2}\right)
+(b+c)\lambda^2(\mu)\log\left(\frac{p^2}{\mu^2}\right)
\label{ampl2}
\ee
From the $\mu$-independence of the amplitude we obtain the ``$\mu$-beta function''
\be
\beta_\lambda\equiv\mu\frac{d}{d\mu}\lambda(\mu)=2(a+b)\lambda^2\ .
\ee
In this way the $\mu$-dependence of the coupling contains a spurious part
(the one proportional to $a$) that does not reflect a momentum dependence
in the amplitude, and misses the momentum dependence of the
term proportional to $c$.
This mismatch has the effect that this definition of running coupling
does not solve the problem of the large logarithms that arises when $p$ becomes large,
which is the main reason for the use of the renormalizaton group in
perturbation theory.
Indeed, if we choose $\mu\approx p$ in order to make the second logarithm
in (\ref{ampl2}) small, the first logarithm will generically be large:
We can rewrite the amplitude as
\be
\cM(p)=\lambda(p)+(a-c)\lambda^2(p)\log\left(\frac{m^2}{p^2}\right)\ .
\ee

The problem is solved by using a different renormalization scheme,
where we absorb the first logarithm of (\ref{ampl2})
in the definition of the renormalized coupling:
$$
\lambda(\mu)\to
\lambda(\mu)-(a-c)\lambda^2(\mu)\log\left(\frac{m^2}{\mu^2}\right)
$$
In this scheme the amplitude reads
\be
\cM(p)=\lambda(\mu)
+(b+c)\lambda^2(\mu)\log\left(\frac{p^2}{\mu^2}\right)\ .
\ee
From the requirement that this be $\mu$-independent,
one gets what we call the physical beta function
\be
\beta_\lambda\equiv\mu\frac{d}{d\mu}\lambda(\mu)=2(b+c)\lambda^2\ .
\ee
Now the $\mu$-dependence of the coupling faithfully tracks the
momentum-dependence of the amplitude,
and with this definition of running the problem of the large logarithms is solved.

When could this type of behavior arise?
It does not arise in the $CP(1)$ model considered in the present paper.
This is due to the absence of IR divergences \cite{David:1980rr}, that implies $a=-c$.
Then, the $\mu$-running and the physical running are the same.

Also, it does not arise in textbook discussions of running couplings
in ordinary renormalizable theories in four dimensions,
with kinetic terms involving two derivatives.
In these cases, there are no IR divergences off-shell,
and tadpoles cannot contribute to the renormalization of marginal couplings.

In \cite{Buccio:2023lzo}, we considered a higher derivative four dimensional model with a shift symmetry, for which there were no lower dimension interactions possible. In this case the high energy running of the interaction again agrees in all regularization schemes.
However, the wave function renormalization constant runs with $\mu$,
but this running is not physical.

A more striking example was given in \cite{Donoghue:2023yjt},
that considered a  four dimensional  nonlinear sigma model with four derivatives (which might be thought of a 4D cousin of the 2D O(N) models with two derivatives).
In this theory the symmetry allows a two derivative addition, i.e.
\beq
{\cal L} = \frac{F^2}{4} Tr \left(\partial_\mu U^\dagger \partial^\mu U\right) - \frac1{4f^2} Tr  \left(\Box U^\dagger \Box U\right) +...
\eeq
 with $U(x) = \exp i \tau\cdot \phi(x)$. These couplings appear in the propagator
 \beq
 i Dq) = \frac{i}{F^2 q^2 - \frac1{f^2}q^4}
 \eeq
 and can be renormalized by considering the self energy. However, because only even powers of the field appear in this theory, the self-energy is just the tadpole diagram at one loop.  This would have an IR divergence if the two derivative Lagrangian were omitted. And because the tadpole diagram does not have any momentum dependence, there is a difference 
between cutoff running or $\mu$ running and physical running~\cite{Donoghue:2023yjt}.

This phenomenon is also found in our treatment of Quadratic Gravity, where if one  considered only the curvature squared terms, there would be infrared divergences, which however are removed by the proper inclusion of the Einstein-Hilbert action 
(that is of first order in the curvature). 
Again tracing $\Lambda$ running or $\mu$ running in Quadratic Gravity differed from the physical running \cite{Buccio:2024hys}. The lesson appears to be that when operators of different dimensions appear in the action bringing in relative dimensionful factors, the running with cutoffs and the physical running with momentum should be expected to be different. 

That Green functions could exhibit different dependence on $\mu$ and on momentum,
when higher derivatives are present, had also been observed in \cite{Ghilencea:2004sq}.

In the early days of quantum field theory, Landau, Pomeranchuk and collaborators studied the running couplings in many theories and concluded that all quantum field theories have Landau poles -- i.e. are not asymptotically free 
\cite{Landau:1955ip,Landau:1958jby,Pomeranchuk:1956zz}. 
This was famously overcome by the proof of asymptotic freedom in Yang-Mills theory, and this is often attributed to the extra degrees of freedom found in the non-Abelian gauge theory. The 2D $CP(1)$ model is another interesting counterexample to the Landau argument. Here there are only scalar degrees of freedom. The logarithms of the energy variables can be found by taking the unitarity cuts in all channels. The analysis using either cutoff regularization or dimensional regularization is consistent with this physical running. The $CP(1)$ model shows that non-gauge theories may also exhibit asymptotic freedom.

 \bigskip

{\it Acknowledgements} --  Equivalent results have been obtained by Kiaee and Monin \cite{monin} and we thank the authors for mutual discussions as their calculations have developed. We also thank Andrei Barvinsky, Arkady Tseytlin, and Misha Shifman for useful  discussions. JFD acknowledges partial support from the U.S. National Science Foundation under grant NSF-PHY-21-12800. GM acknowledges partial support from Conselho Nacional de Desenvolvimento Cient\'ifico e Tecnol\'ogico - CNPq under grant 317548/2021-2, Funda\c{c}\~ao Carlos Chagas Filho de Amparo \`a Pesquisa do Estado de S\~ao Paulo - FAPESP under grant 2023/06508-8 and Funda\c{c}\~ao Carlos Chagas Filho de Amparo \`a Pesquisa do Estado do Rio de Janeiro - FAPERJ under grant E-26/201.142/2022.

\end{document}